\documentclass[conference]{IEEEtran}
\IEEEoverridecommandlockouts
\usepackage{cite}
\usepackage{amsmath, amssymb, amsfonts}
\usepackage{algorithm, algorithmic}
\usepackage{graphicx}
\usepackage{textcomp}
\usepackage{xcolor}
\usepackage{url}
\def\BibTeX{{\rm B\kern-.05em{\sc i\kern-.025em b}\kern-.08em
		T\kern-.1667em\lower.7ex\hbox{E}\kern-.125emX}}

\begin{document}
	\title{BlockFW - Towards Blockchain-based Rule-Sharing Firewall \thanks{Publication with Open Access in The 16th International Conference on Emerging Security Information, Systems and Technologies (SECURWARE), pp. 70-75, IARIA 2022.}}
	
	\author{
		\IEEEauthorblockN{Wei-Yang Chiu and Weizhi Meng}
		\IEEEauthorblockA{SPTAGE Lab, Department of Applied Mathematics and Computer Science, \\Technical University of Denmark, Denmark}
		}

	\maketitle
	%However, it seems that it failed partially due to lack of validating and monitoring.
	\begin{abstract}
Central-managed security mechanisms are often utilized in many organizations, but such server is also a security breaking point. This is because the server has the authority for all nodes that share the security protection. Hence if the attackers successfully tamper the server, the organization will be in trouble. Also, the settings and policies saved on the server are usually not cryptographically secured and ensured with hash. Thus, changing the settings from alternative way is feasible, without causing the security solution to raise any alarms. To mitigate these issues, in this work, we develop BlockFW -- a blockchain-based rule sharing firewall to create a managed security mechanism, which provides validation and monitoring from multiple nodes. For BlockFW, all occurred transactions are cryptographically protected to ensure its integrity, making tampering attempts in utmost challenging for attackers. In the evaluation, we explore the performance of BlockFW under several adversarial conditions and demonstrate its effectiveness.
	\end{abstract}


\section{Introduction}
It is difficult to overlook security policies over large networks for network administrators. When attacks occurred from either internal or external network, it can be quite challenging for them to quickly take measures and deploy new policies~\cite{Almutairi2022,Meng2011}. For example, performing penetration test toward multiple servers in a network can be quite simple~\cite{Rak2022}, such as setting up scripts for automating the attack. However, it is quite an opposite situation for network administrators, since collecting information and deploying security solutions need to be done one-by-one. This is very time-consuming and labor-costly compared to performing an attack. To overcome this unfair situation, commercialized central-managed security solutions are provided by many security providers. These products give administrators a dashboard or a cockpit, making it easier to overview situations in the network. That is, information can be collected, and policies can be deployed at one-stop.
	
However, what these solutions are offering can also become a security breaking point of the system~\cite{Meng2014_cose}. All endpoints, by default, must trust the decision and command coming from the central server of the security solution. If the management server is compromised, it can become a huge loophole of the security status in an organization~\cite{Trope2016}. For example, attackers can command all security solutions deactivated in order to reveal further exploits of the internal network.
	
	\begin{figure}
		\centering
		\includegraphics[width=\linewidth]{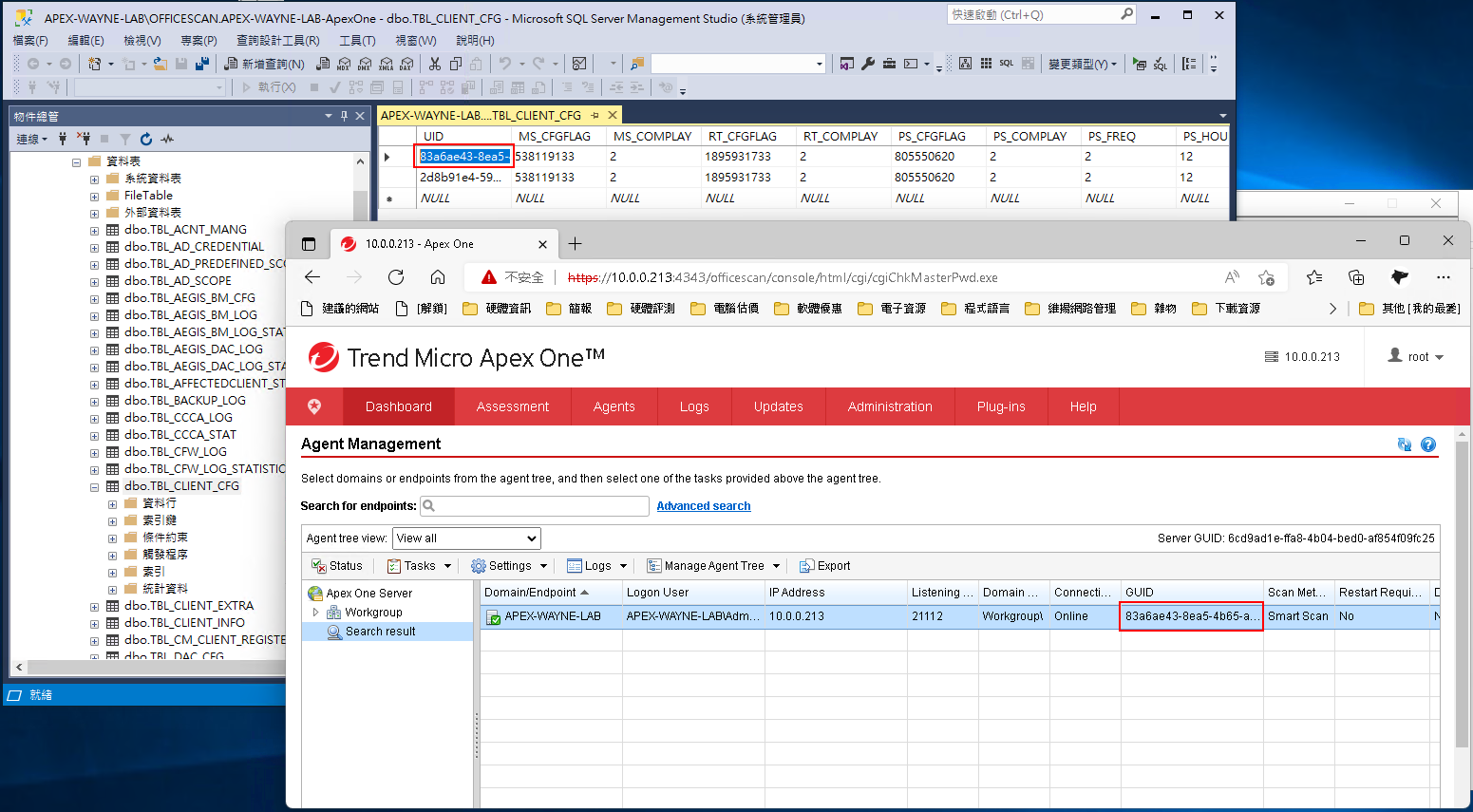}
		\caption{A Centralized Security Solution with Database in Mutable Storage}
		\label{fig:intro:secDatabase}
\vspace{-2mm}
	\end{figure}
	
\begin{figure}
 		\centering
 		\includegraphics[width=\linewidth]{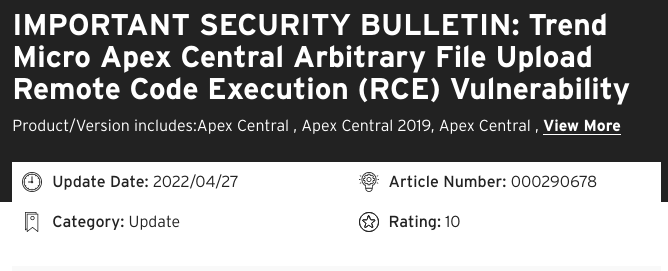}
 		\caption{Security Solution Vulnerability}
 		\label{fig:intro:secVul}
 \vspace{-2mm}
 	\end{figure}

Fig.~\ref{fig:intro:secDatabase} shows an example of a central-managed security solution with its settings stored in a mutable database. We can perform some value changes, not through the security solution's management console, but through the database console. Then we notice the existence of toolkit that can directly access the offline database file, without any restrictions from the configured database management system. Although the attackers could not obtain the management console's access credential, they have a good chance to change the security solution's settings through several alternative methods, which can be considered as unauthorized changes for the security solution. In this case, although attackers may not be able to find the exploit to the security solution itself, they can still affect the security policies via different vulnerabilities on the server that holds the centralized management of the security solutions, as shown in Fig.~\ref{fig:intro:secVul}.
 	
	The above potential threat creates the need of having a second pair of eyes to closely monitor the management server itself, making organizations with centralized security solutions more insecure. However, as we closely inspect the example case we are studying, we can see that the issue itself is more related to the underlying database. In other words, changes of a security policy can be made through alternative routes that are outside the designed workflow, which requires the validation and the monitoring from others in the environment.
	
\textbf{Motivation.} As blockchain becomes a constantly discussed topic recently, several of its characters can tackle the issues of central-managed security solutions~\cite{Li2022_tis,Meng2021_infu}. They are the immutability of occurred events, and evidence of transaction events is cryptographically strengthened so that data integrity will become extremely challenging to compromise, and the underlying consensus algorithm will be able to follow one version of the data with their recognition. Further, blockchain requires its participants to hold a partial or full copy of the network transaction log, called \emph{ledger}. Transactions are collected and validated by network maintainers, such characters or equivalent may have different names in different platforms, before being cryptographically sealed into a basic storage unit, named \emph{block}. Generated block contains the cryptographical proof (e.g., hashes) of the previous blocks. This creates a strengthened chain-like storage structure, which is challenging to break~\cite{Chiu2021_ppna,Chiu2021_ijsa}. For attackers that would like to alter the previously existed records, it will be extremely time-taking, making such operation infeasible.

If attackers deliberately change the database records by editing it forcefully, it will result in either the node being ditched out of the network due to tremendous differences, or the tampered database records will be restored from other nodes~\cite{Mukta2022}. Both situations are not favorable to the attackers.
	
\textbf{Contributions.} In this paper, our main goal is to deploy a proof-of-concept of centralized security management on top of blockchain, in order to showcase the feasibility and resilience of such system under cyber-attacks. In particular, we develop BlockFW -- a blockchain-based rule-sharing firewall, and investigate its performance under adversarial conditions. The results indicate its capability of lowering the cost of operating a security solution. %though we limited functionality, already shows its capability, and even in some situation, lowering the cost of operating a security solution.

The rest of this paper is organized as follows. Section~\ref{sec:2} introduces the background and related work. Section~\ref{sec:3} details the design of BlockFW including the requirements and major components. Section~\ref{sec:4} presents the performance evaluation under some adversarial scenarios. Section~\ref{sec:5} concludes the work with future work..

	%We will shortly introduce what Blockchain is, the evolution of security solutions, and the attempts of combining the two. Later, we will dive into the system design and evaluation. Finally, we will end the paper with discussion and future works.

\section{Background and Related Work} \label{sec:2}
This section introduces the background on blockchain and consensus algorithm, and discusses the related studies.

\subsection{Blockchain}

Blockchain, by its design and practice, is considered as a kind of decentralized ledger technology (DLT)~\cite{Chiu2021_ijsa,Li2022_tis}. A block is the basic storing unit in the blockchain, which can be formed in a periodic way including the collected transactions within a time period. A consensus algorithm is applied in the network to allow everyone validating the blocks and to reach an agreement on the block version. Basically, consensus algorithm will select a sealer to seal the latest formed block with strong cryptography. The block is then distributed to all network participants for updating their local copies. %This chained-like data storing structure is where Blockchain gets its name.

To ensure the unification of the decentralized database is the primary designing goal of a consensus algorithm. Below are two typical algorithms.

\paragraph{Proof-of-Work (PoW)}
A PoW-based system will generate a challenging computational problem, in which a difficulty control mechanism is involved. The level of difficulty can be adjusted according to the system's requirements. The participant who first solves the problem will win the turn. %(See Figure \ref{powstep}) We can shortly conclude that PoW is responsible for ensuring that each block can only have one sealer and providing the sealing hash for the block.

%PoW, rather than being a widely deployed consensus algorithm for the blockchain platform, is initially designed for DoS attack protection. A requester that would like to obtain service must solve the problem from that particular. Otherwise, the request from the requester will be considered invalid. The design creates a barrier for those sending requests nonetheless. For attackers who like to reach the server, it creates a tremendous computing burden for attackers to develop an effective attack. This idea is presented in 1993 by Dwork and Naor~\cite{b29}. Later, the term of PoW is coined by Jakobsson and Jules in 1999~\cite{b30}.

Being the first consensus algorithm in Bitcoin~\cite{b34} with the easy-to-understand design philosophy, PoW indeed dominates the market of cryptocurrencies. However, with the network participants increasing,  many new challenges can be caused, i.e., the tremendous waste of computational power on completing transactions. Profitable mining activities may encourage the forming of mining pools.  The concentration of computing power leads to the threat of 51\% attack~\cite{b32}. That is, when a particular group owns 51\% or more computational power of the whole network, it has unsurpassed domination on manipulating future records~\cite{b31}. %By denying and refuse to serve the unwanted request, the group can forcefully insert malicious records. Although the attack may sound ridiculous, it no denies that some platform has the potential vulnerability toward 51\% attack~\cite{b33}.

\paragraph{Proof-of-Stake (PoS)}
As a possible solution to complement PoW consensus algorithm, PoS chooses sealers by rounds of selection rather than computing competitions. More specifically, PoS asks participants to take some of their assets (or coins) to join the election. The system chooses the preferable stake by conditions. The selected stake's owner wins the turn~\cite{Meng2018_access}. The criteria of how the system decides the preferable stake is crucial. For example, setting the criteria as preferring a larger stake may cause monopoly. For this issue, \emph{coin-age} that measures a coin's stagnation in an account is considered as a promising solution~\cite{Baldimtsi2020}.

PoS provides a more power-efficient method of reaching consensus and providing more fairness of sealer selection toward the participant with less computational power. However, it does not prevent the 51\% attack. Though PoS does not suffer from the monopoly of computational power, it may suffer from the monopoly of wealth. As opposite to 51\% of computational power, 51\% of the wealth can provide unsurpassed advantages on winning the stake~\cite{Chiu2021_acisp}.

\subsection{Related Work}
The application of blockchain technology in developing a firewall is not new. In the literature, Steichen \emph{et al.}~\cite{Steichen2017} introduced ChainGuard, which could use SDN functionalities to filter network traffic for blockchain-based applications. Their system required that all traffic to the blockchain nodes has to be forwarded by at least one of the switches controlled by ChainGuard. Li \emph{et al.}~\cite{Li2022_jsa} then developed a blockchain-based filtration mechanism (similar to firewall) with collaborative intrusion detection to help protect the security of IoT networks by refining unexpected events. It is found that though some ideas have been proposed on blockchain-based firewall, they have not been widely implemented. This motivates our work to implement a prototype of blockchain-based firewall and examine its performance in a practical setup.

Many research studies are focusing on the combination of blockchain technology with intrusion detection. For instance, Meng \emph{et al.}~\cite{Meng2020_ijis} designed a blockchain-based approach to help enhance the robustness of challenge-based intrusion detection against advanced insider attacks, where a trusted node may suddenly become malicious. Li \emph{et al.}~\cite{Li2022_tis} introduced BlockCSDN, a framework of blockchain-based collaborative intrusion detection for Software Defined Networking (SDN). A similar scheme was also proposed by Meng \emph{et al.}~\cite{Meng2021_infu}, which used blockchain to enhance the robustness of trust management. Some more relevant studies can refer to surveys~\cite{Al-Kadi2020,Li2022_tis,Li2022_comst}.

\section{BlockFW - A Blockchain-based Rule-Sharing Firewall} \label{sec:3}
	
This section introduces how our proposed blockchain-based rule-sharing firewall works. At first, we briefly describe how to choose and decide a blockchain platform for our case. Then we present the high-level architecture of our system including the major software components.% and performance evaluation. Finally, we finish the topic with conclusion and discussion.
	
\subsection{The Requirement for underlying Blockchain Platform}
	Although different blockchain platforms share similar concepts, the underlying implementation differences provide the platforms with various advantages separately. Not all platforms can become the data storage of our system. For our purposes and goals, we consider a suitable platform that should have the following characteristics:
	
	\begin{itemize}
		\item{\verb|Semi-Dynamic Network|:} Servers may be added or removed according to the changes or expands in services. In the trend of X-as-a-Service, cloud, and virtualization, the action of adding or removing service entities can be dynamic. Though being dynamic, there are differences from the public network: authentication is mandatory. Nodes in the network cannot join or leave the network autonomously, authorization entity or authorized personnel must get involved and approve the operation. This specific characteristic creates a semi-dynamic all-known-nodes network. Furthermore, since all network nodes are responsible toward different tasks and may potentially be vulnerable in different ways, we have to assume that part of the network may become malicious. Hence the network we are trying to deploy must be Byzantine-resistible.
		
		\item{\verb|Stable Connection|:} Since servers are regarded as critical infrastructure in IT-enabled businesses, they are usually either connected through the internal network, or the connections can be ensured by telecom SLA with the company. Compared with the wide area network, it has less flickering or instability issues. We consider that it can accept having a blockchain-platform with higher counts of exchanged messages during communication.
		
		\item{\verb|Timing-Sensitive|:} When attacks occurred, we definitely expect that the traffic can be blocked as soon as possible when being a network administrator. However, even deploying security policies through many centralized security solutions may take a while to reach every client. Although it is unreasonable to have everything responded at instant, the actions have been taken will reach and execute by clients eventually. While the time consumption should be in a reasonable length from the command being given to the action being taken. Thus a blockchain system that completes transactions in an estimable time is important.
		
	\end{itemize}
	
Based on the above characteristics, we figure out that our BlockFW platform needs to be Byzantine tolerable with stable transaction speed, in which these requirements are usually satisfied in a private blockchain.
	
In this work, we decide to implement the system based upon the DevLeChain platform~\cite{devlechain} -- a blockchain development environment, which can be used to quickly and easily set up a desired environment~\cite{Chiu2022_blockchain}. In addition, it supports multiple different blockchain platforms. Hence, we can easily switch between platforms to observe the differences.

	\begin{figure}
		\centering
		\includegraphics[width=\linewidth]{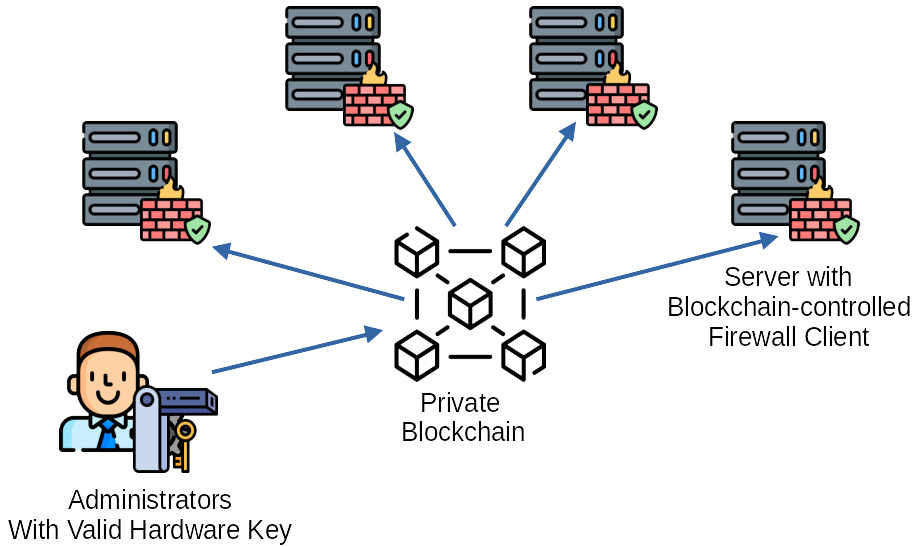}
		\caption{The Overview Structure of BlockFW}
		\label{fig:blockfw:sysOvervivew}
\vspace{-3mm}
	\end{figure}

\subsection{The System Overview}
	
As shown in Fig. \ref{fig:blockfw:sysOvervivew}, BlockFW features a simple and straightforward system structure, which consists of two major roles and three major pieces of software.
	
	\begin{figure*}[t]
		\centering
		\includegraphics[width=.6\linewidth]{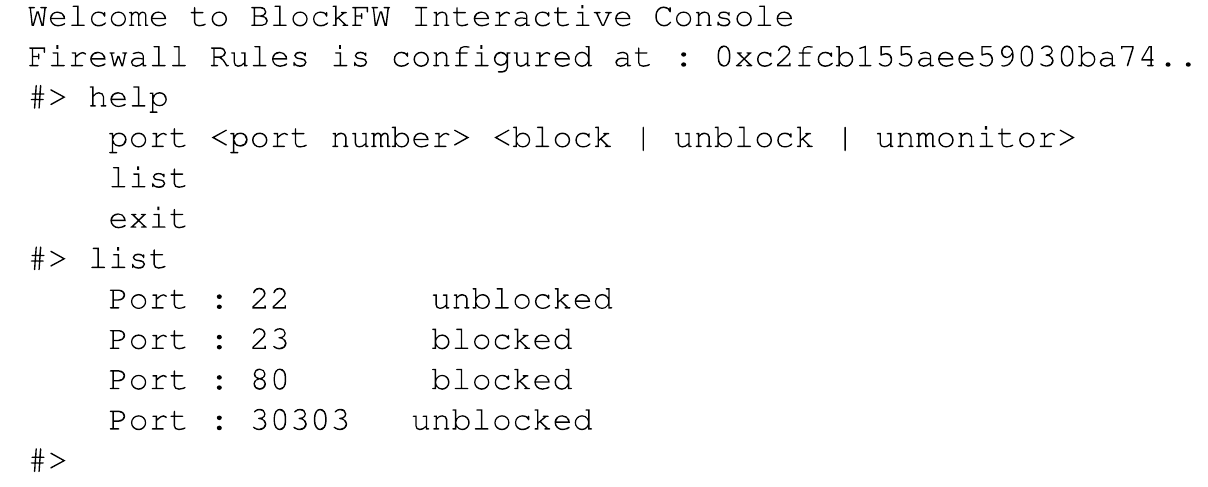}
		\caption{The BlockFW Management Console}
		\label{fig:blockfw:mgmtConsole}
\vspace{-3mm}
	\end{figure*}

	The two roles are:
	\begin{itemize}
		\item{\verb|Administrators|:} They have the permission to set and alter firewall rules to the system. Each administrator will be given a hardware key that has been registered into the system. Existing administrators can set other keys as administrators. The hardware key is regarded as the wallet file of the administrator when interacting with the blockchain.
		\item{\verb|Clients|:} These are endpoints that listen and monitor the given rules on the blockchain. They are installed with firewall software, which can act according to the rules on the blockchain.
	\end{itemize}

\begin{figure}[t]
		\centering
		\includegraphics[width=.85\linewidth]{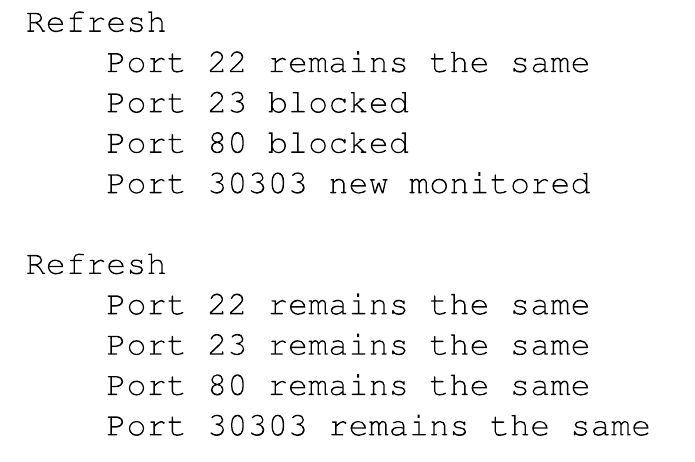}
		\caption{The BlockFW Management Console}
		\label{fig:blockfw:fwCmd}
\vspace{-3mm}
	\end{figure}

	The three major software components are:
	\begin{itemize}
		\item{\verb|Management Console|:} The console is a command-line interface for administrators to add new firewall rules or manage existing firewall rules, as depicted in Fig. \ref{fig:blockfw:mgmtConsole}. It requires the administrator's hardware key to function correctly. If a non-registered hardware key is provided, any command given to the management console will fail. This is because the system's backend smart contract is enforced with Access Control List, which contains the public-key-derived wallet addresses. Any non-registered key  will result in transactions that are unacceptable to the smart contract, as it cannot be validated.
		
		\item{\verb|Firewall-Commander|:} The firewall-commander is the middleware between the blockchain and the system. It monitors the blockchain for any changes periodically. If the current firewall state is different from what the blockchain has stated, it will synchronize the rules in local system firewall, as shown in Fig. \ref{fig:blockfw:fwCmd}.
		
		\item{\verb|Blockchain|:} The blockchain is acted as the decentralized database among clients and administrators.
	\end{itemize}

\section{Performance Evaluation} \label{sec:4}
In this section, we present the environmental setup and evaluate the system under different adversarial scenarios.

\subsection{System Configuration}
	
To test how effective the proposed system is, we configured \textbf{three nodes} with client installed, \textbf{one administrator node} with hardware key, and \textbf{one attacking node} toward the network. Each node is given and configured with the information listed in Table \ref{tab:eval:sysConf}.
	
	\begin{table}
		\caption{Environmental Platfrom}
		\label{tab:eval:sysConf}
		\begin{tabular}{|p{1cm}|p{2.4cm}|p{1.2cm}|p{2.2cm}|}
			\multicolumn{2}{c}{VM Reousrces} & \multicolumn{2}{c}{Software}\\
			\hline
			Item & Config. & Item & Version \\
			\hline
			CPU & Intel Xeon W-2133 @ 3.6GHz x2 & Hypervisor & vmware ESXi 7.0 U3d \\
			\hline
			Memory & 4GB ECC DDR4-2666 & Guest OS & mxLinux 21 \\
			\hline
			Storage & 48GB HDD & Blockchain Platform & Ethereum 1.10.18 \\
			\hline
			Network & vmware vSwitch 1G & Contract Platform & EVM \\
			\hline
		\end{tabular}
\vspace{-2mm}
	\end{table}

\begin{figure}
		\centering
		\includegraphics[width=\linewidth]{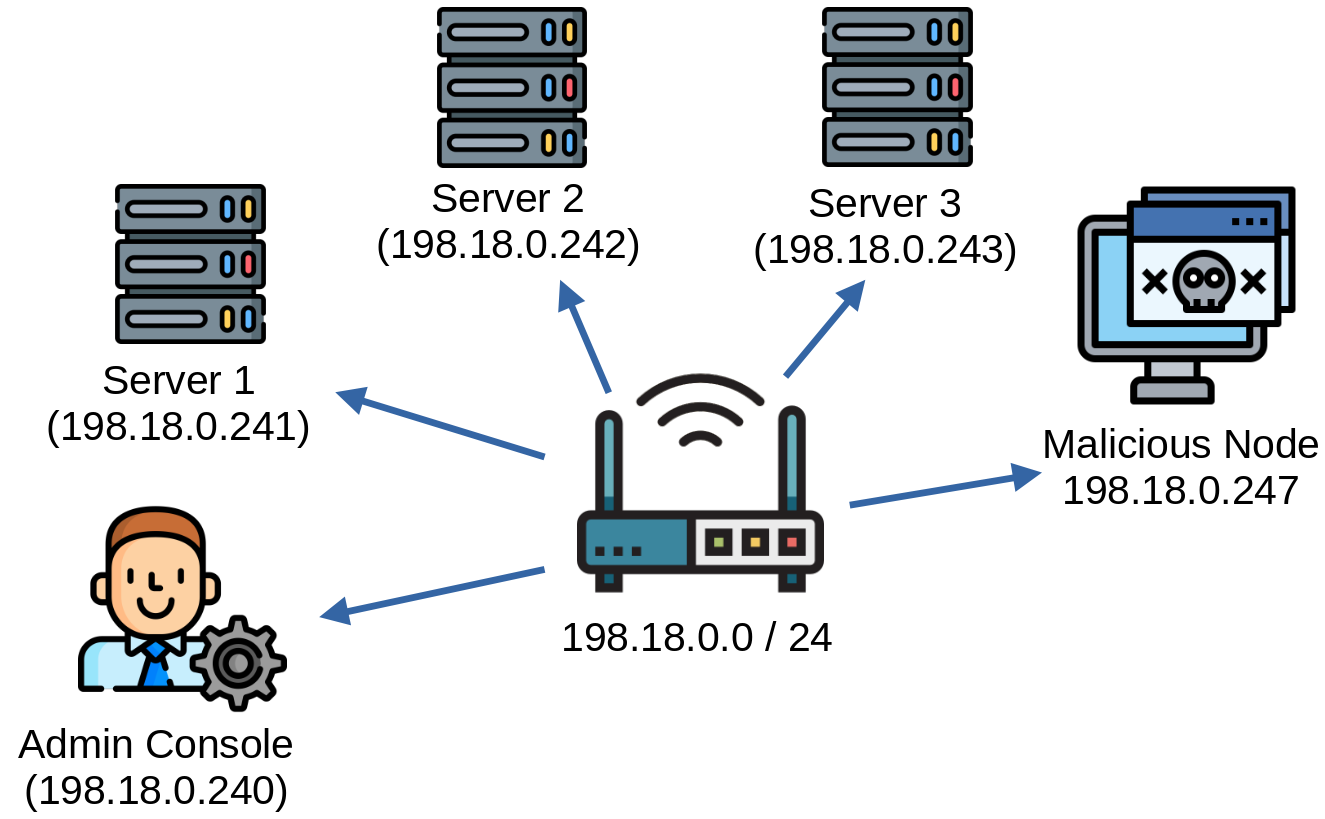}
		\caption{The network configuration for the testing environment}
		\label{fig:eval:network}
\vspace{-3mm}
	\end{figure}

	For concise and clear demonstration, we set up all entities under the same network, as illustrated in Fig. \ref{fig:eval:network}. Servers are running three common services: the SSH (Port 22), the Telnet (Port 23), and the HTTP (Port 80).
	
\subsection{Experiment-1: Attacking toward a Group of Servers}
\begin{figure}[h]
		\centering
		\includegraphics[width=.45\linewidth]{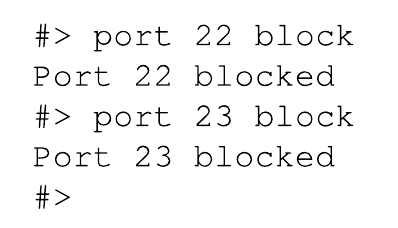}
		\caption{Blocking Port 22 and Port 23 through the Console}
		\label{fig:eval:PortBlock}
\vspace{-3mm}
	\end{figure}

In this test, we assume that malicious node can brute-force the SSH and Telnet, while sending invalid HTTP packet to the web server.
If any centralized security solution has not been implemented, then administrators have to do it one by one. In the comparison, our blockchain-based solution can complete this task more quickly. For example, the administrators can use the following commands via the management console, as demonstrated in Fig. \ref{fig:eval:PortBlock}.

\begin{figure} [t]
		\centering
		\includegraphics[width=.45\linewidth]{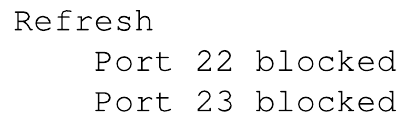}
		\caption{Client updating firewall rules}
		\label{fig:eval:clientUpdate}
\vspace{-3mm}
	\end{figure}

	In particulary, we configured the Firewall-Commander to refresh the rules every 5 seconds, as Clique consensus algorithm can finish packaging and generate blocks very quickly, as shown in Fig.~\ref{fig:eval:clientUpdate}. It is guaranteed that the Firewall-Commander can reach the updated rules within the refreshing period.

\begin{figure} [h]
		\centering
		\includegraphics[width=\linewidth]{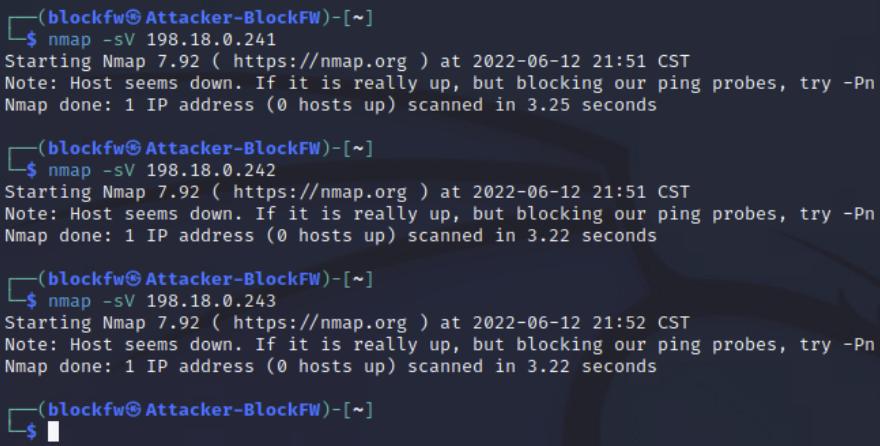}
		\caption{The NMAP scanning result of the server}
		\label{fig:eval:nmapScan}
\vspace{-3mm}
	\end{figure}

	After updating the firewall rules, the attack could be instantly stopped as shown in Figure \ref{fig:eval:nmapScan}. The attacker cannot performs either SSH or Telnet to the protected servers.
	
\subsection{Experiment-2: When the Network is under Stressed}

Many centralized security solutions can be often affected under the Denial-of-Service (DoS) attack. If the traffic flow was stressed out the centralized management server, it becomes difficult for clients to send or receive heartbeat toward and from the server. Hence, the deployment of security rules may become challenging.
	
	Although blockchain is, theoretically, not affected much from DoS attacks toward single node, we still have to know how much it may affect the system. Consensus algorithms, especially those for private chains, have intensive message-exchange protocols. In this case, if the message could not be effectively exchanged, it will affect the rule deployment.
	
\begin{figure}
		\centering
		\includegraphics[height=.9\linewidth]{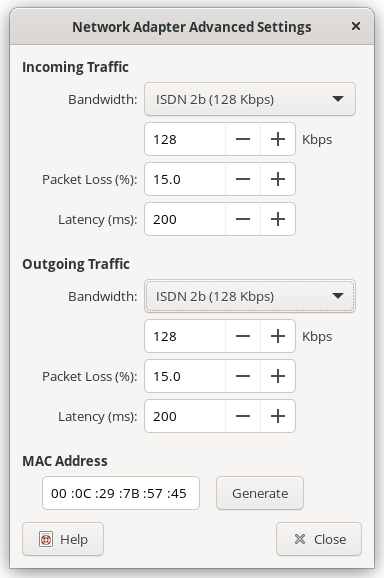}
		\caption{Creating an unreliable network}
		\label{fig:eval:networkSim}
\vspace{-2mm}
	\end{figure}

	However, it is difficult to perform the experiment by really stressing the nodes with loads, as they are all on the same machine, and even the network switch is emulated. However, it does not mean that we could not emulate the environment through different ways. In this experiment, we deliberately configured the vSwitch~\cite{b35} to emulate an unreliable network environment, as shown in Fig. \ref{fig:eval:networkSim}. We configured the network with the following parameters:

	\begin{itemize}
		\item{\verb|Bandwidth|:} 128 kbps Full-Duplex
		\item{\verb|Packet Loss|:} 15.0\%
		\item{\verb|Latency|:} 200 ms
	\end{itemize}
	
\begin{figure}
		\centering
		\includegraphics[width=\linewidth]{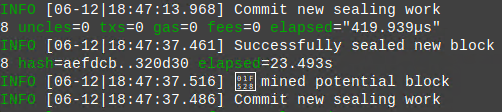}
		\caption{The mining output from the console}
		\label{fig:eval:miningOut}
\vspace{-3mm}
	\end{figure}

As shown in Fig.~\ref{fig:eval:miningOut}, the time of generating a new block instantly bumped up to around 23 seconds. Other nodes that do not join the mining took another 2-3 seconds to receive the new block. On the Firewall-Commander console, it took around 30 seconds on average to complete the deployment of new firewall rules.
	
Overall, it is found that our BlockFW system can still work under a stressed network, while the speed of making a policy may slow down. On the positive side, though it is becoming slower, the policy is still reachable to the endpoint.
	
\subsection{Experiment-3: When a Server is Tampered}
	
As long as the administrator's hardware key is removed from the system, the smart contract on the blockchain cannot be altered. However, we still tried to deliberately corrupt the ledger copy in one of the servers, in order to investigate how the system will react under this condition.
	
\begin{figure}
		\centering
		\includegraphics[width=\linewidth]{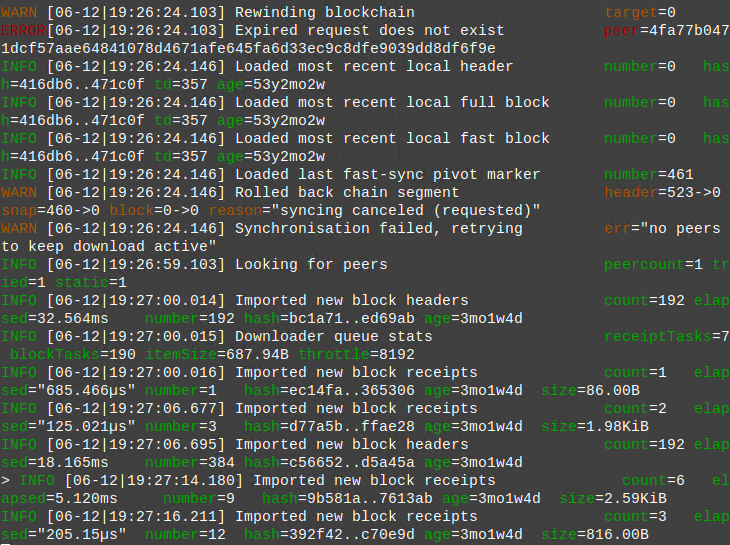}
		\caption{Blockchain Synchronization Triggered}
		\label{fig:eval:syncTrig}
\vspace{-3mm}
	\end{figure}

More specifically, we deliberately blank out one of the blockchain database files, and see how the system reacts. As shown in Fig. \ref{fig:eval:syncTrig}, it is found that the blockchain client detected these anomalies in the local ledger, and immediately started to sync with other nodes.%, as indicated in Figure \ref{fig:eval:syncTrig}.
	
	In conclusion, although an attacker can deliberately tamper a local ledger copy, the blockchain client will instantly notice the anomalies, start downloading chain data from other nodes and replacing the corrupted local copy. In this case, our BlockFW can be more robust than a centralized security solution, if the server is under attack.
	
\section{Conclusion and Future Work} \label{sec:5}
	
In this paper, we developed a blockchain-based rule-sharing firewall (called BlockFW) that can offer validation and monitoring among multiple nodes. In the evaluation, we tested BlockFW in several harsh network conditions and investigated whether it can perform better than a traditional central-managed security solution. Based on the results, it is found that our blockchain-based solution can continue to serve correctly under a stressful network condition. Also, as no central server exists in our system, there is no use for attackers to stress out one of the servers to crash the system. We further demonstrated the adversarial scenario when attackers tried to modify the policies by directly editing the blockchain storage file on one node, and identified that our system could recover itself from other reachable nodes, making the attacker's tampering trial unsuccessful. These provide a good evidence that making blockchain as the underlying database for the security solution is viable with particular advantages.
	
However, the BlockFW system we are developing requires some further improvements. On functionality phase, the implementation is less than a traditional firewall has, in which we are actively developing a new version to overcome this issue. Another important topic that we have not discussed is whether BlockFW can handle a large network the same as the current central-managed security solutions. This is because permission-based blockchain has to utilize voting-based consensus algorithms that require to exchange many messages to reach consensus compared with a traditional lottery-based consensus algorithm (e.g., PoW / PoS). Too many nodes may result in slowdown and a waste of network resources. Thus, the scalability issues are always important when developing a blockchain-based solution. %Although issues under same category for current central-managed solutions happen, we need to find out whether a decentralized solution can work as the same performance as what we have right now.
	
	%Currently, we are completing and extending the functionalities further, and considering including Intrusion Detection System functionalities in the future.

\section*{Acknowledgment}
This work was funded by the European Union H2020 DataVaults project with GA Number 871755. The source code of BlockFW is available at SPTAGE Lab: \url{https://nopkirouter1.compute.dtu.dk/project/blockfw.zip}.

\end{document}